\documentclass[aps,pre,groupedaddress,showpacs,twocolumn]{revtex4-1}
\usepackage[dvipsnames]{xcolor}
\usepackage{hyperref}
\usepackage{graphicx}
\usepackage{amsmath}
\usepackage{amssymb}
\usepackage[export]{adjustbox}
\usepackage{subcaption}
\begin{document}

\title{Multi-phase autoresonant excitations in the Korteweg-de-Vries system}
\author{L. Friedland$^{1}$}

\author{A. G. Shagalov$^{2}$}
\address{$^{1}$Racah Institute of Physics, The Hebrew University, Jerusalem
91904, Israel.\\
$^{2}$Institute of Metal Physics, Ekaterinburg 620990, Russian
Federation.
}
\begin{abstract}
Autoresonant (continuously phase-locked) two-phase waves of the Korteweg-de-Vries equation are excited and controlled using a two-component, small amplitude, chirped frequency driving. These solutions are analyzed in the weakly nonlinear regime. The theory is based on Whitham's averaged variational principle. The problem is reduced to a fully separated two degrees of freedom dynamical problem. This separation allows for simple derivation of the autoresonant thresholds on the driving wave amplitudes. We also excite more complex 4-phase autoresonant waves. The inverse scattering analysis of this case indicates again the separation of all four degrees of freedom in the associated weakly nonlinear dynamical problem.
\end{abstract}

\maketitle

\section{Introduction}
Excitation and control of large-amplitude nonlinear waves still remains one of the challenging problems of nonlinear physics. Typically these waves are described by nonlinear partial differential equations, where formation of even well-known large-amplitude solutions requires starting from not easily realizable initial conditions. Nevertheless, there exists one approach for excitation of such waves by starting from trivial initial conditions. This approach is based on the phenomenon of autoresonance \cite{Wiki}. Wave autoresonance is a continuing (automatic) phase-locking of a driven wave to driving perturbations, yielding non-trivial, large-amplitude solutions. The control of the excited wave is achieved simply by time variation (chirping) of the driving frequency and does not require any feedback. Examples of this approach include excitation of solitons \cite{Aranson}, Faraday waves \cite{Ben-David},  plasma waves \cite{Luo}, and more. Furthermore, autoresonance allowed the formation and control of more complex multiphase waves of form $f=f (\theta_1,\theta_2,...,\theta_N)$, where $\theta_i=k_ix-\omega_it$. In the past, these multiphase autoresonant waves have been studied in the context of the nonlinear Schrodinger equation (NLSE) \cite{Lazar1}, the sine-Gordon equation \cite{Shagalov}, and the Kortweg-de-Vries (KdV) equation \cite{Lazar2}. The standard analysis of these solutions involved non-trivial inverse scattering theory (IST) \cite{Novikov}.

In this work, we analyze the initial stage of autoresonant excitation of multiphase solutions of the KdV equation, including the problem of thresholds on the driving amplitudes, by applying a simpler quasi-linear theory and avoiding the need of using the IST. Recently, a similar approach has been applied in studying the formation of space-time quasicrystals in Bose-Einstein condensates \cite{Lazar3} and plasma waves \cite{Munirov}.
Our presentation will be as follows. In Section II, we illustrate the formation of multiphase KdV waves and discuss the phenomenon of the autoresonance threshold on the driving amplitudes in numerical simulations. Section III describes our weakly nonlinear Lagrangian approach based on Whitham's averaged variational principle \cite{Whitham}. In Section IV, we focus on two-phase excitations and reduce the problem to two degrees of freedom dynamical system in action-angle (AA) variables. We present further illustration of the validity of our threshold theory based on the separation of degrees of freedom in the quasilinear approximation using the IST approach in excitation of two- and four-phase autoresonant waves in Sec. V. Finally, Section VI presents our conclusions.

\section{Two-phase waves in numerical simulations}
\begin{figure}[bp]
\includegraphics[width=0.5\textwidth, height=0.30\textheight, left]{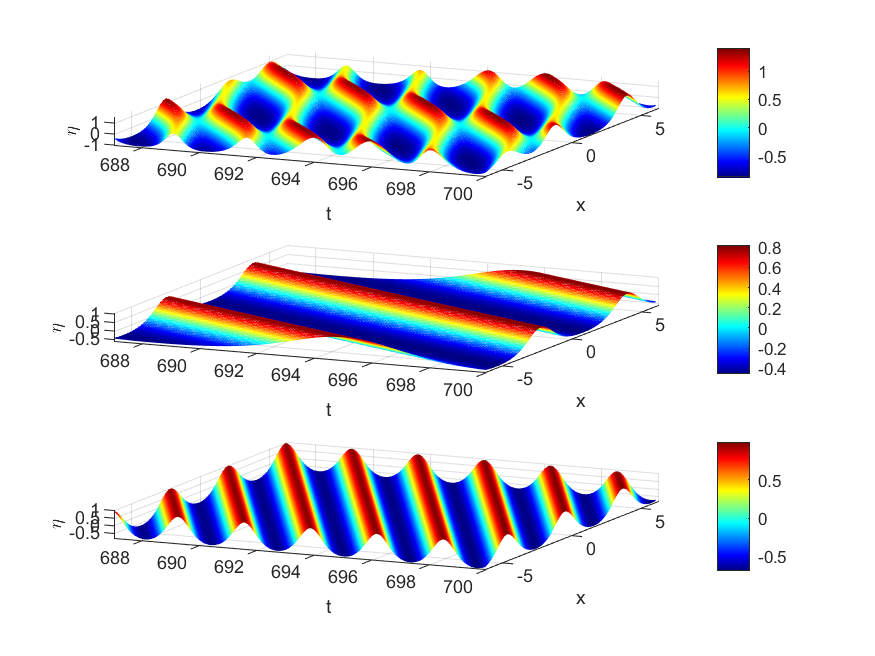}
\caption{The upper panel shows autoresonant two-phase KdV wave excited using above-threshold driving amplitudes $\epsilon_{1,2}=1.1\epsilon_{th}^{1,2}.$ 
The two lower panels show autoresonant single phase (traveling) nonlinear waves when only one of the drivings is applied.}
\label{fig1}
\end{figure}
\begin{figure}[tbh]
\includegraphics[width=0.53\textwidth, height=0.23\textheight, left]{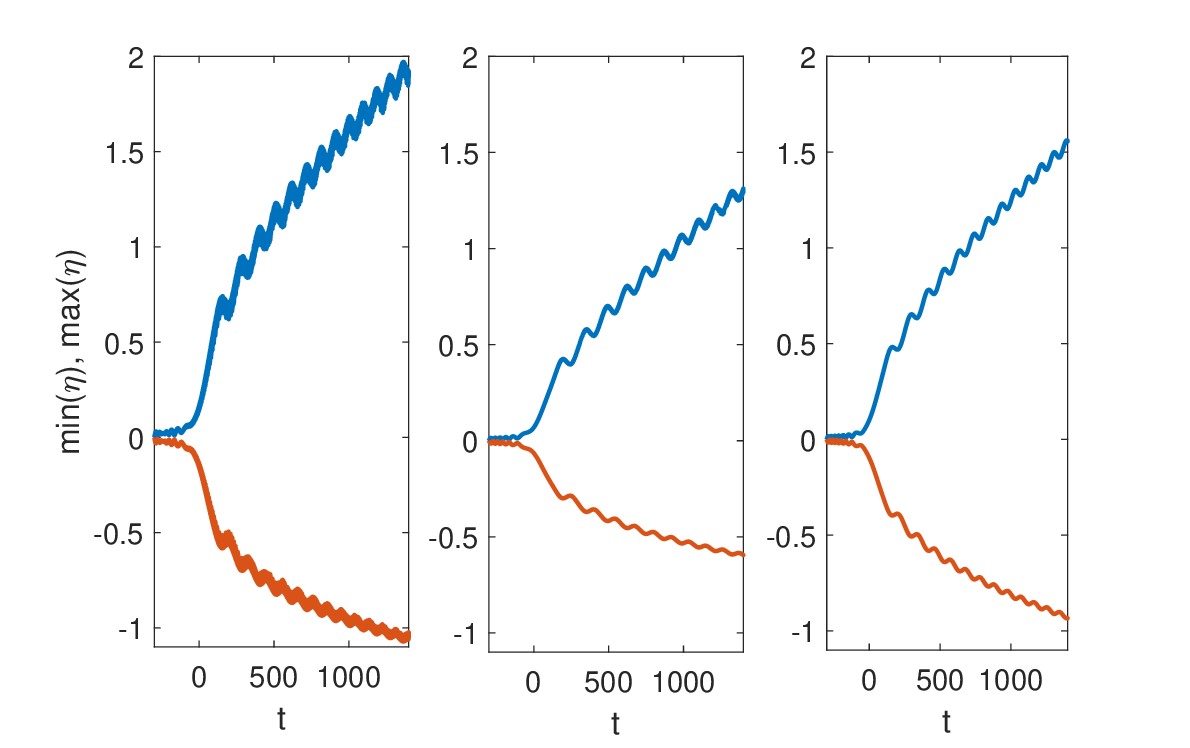}
\caption{The evolution of the maxima and minima (over $x$) of the two- and one-phase autoresonant waves. The left, middle, and right panels correspond to the upper, middle, and the lowest panels in Fig.1, respectively.}
\label{fig2}
\end{figure}
We consider dimensionless driven KdV equation
\begin{equation}
\eta _{t}+6\eta \eta _{x}+\eta _{xxx}=\epsilon _{1}\cos \theta
_{d1}+\epsilon _{2}\cos \theta _{d2},  \label{1}
\end{equation}%
where $\theta _{di}=k_{i}x-\int \omega _{di}(t)dt$ and $\omega _{di}(t)$ are
slowly varying driving frequencies. We solve this system numerically for the two driving components in a time interval $-300<t<700$ using the method of Ref. \cite{Lazar2}. Figure 1 shows the results of our simulations with zero initial condition (at $t=-300$) in the final stage $480<t<700$, using driving wave numbers $k_1=1$, $k_2=1.5$, slowly chirping driving frequencies $\omega_{di}=-k_i^3+\alpha_it$ ($\alpha_{1,2}=0.0008,0.001$), and three sets of driving amplitudes $\epsilon_{1,2}$. In the upper panel in Fig. 1, both driving components are present, while in the two lower panels one of these components vanishes ($\epsilon_{2}=0$ or $\epsilon_{1}=0$, respectively). The nonzero driving amplitudes in all these examples are $10\%$ above the threshold values $\epsilon_{th}^i$
for autoresonant excitation [see Eq. \eqref{28} derived in Sec. IV].
\begin{figure}[tbh]
\includegraphics[width=0.47\textwidth, height=0.25\textheight, left]{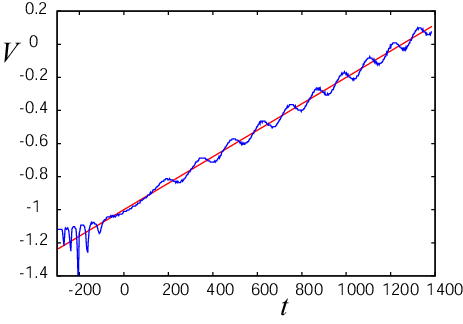}
\caption{The phase velocity of the driven single-phase autoresonant wave (in blue) and that of the driving wave (in red)}.
\label{fig3}
\end{figure}
We also show in Fig. 2 the maxima and minima over $x$ of the excited waves as functions of time in the full time interval of integration in the same three examples as in Fig. 1. One can see continuously growing amplitude solutions and superimposed slow modulations. This growth despite variation of the driving frequency and the modulations are characteristic of a continuing resonance in the driven system \cite{Wiki}. A loss of phase-locking would lead to saturation of the amplitude. The color maps of the excited solutions in the lower two panels in Fig. 1 illustrate large-amplitude traveling autoresonant cnoidal waves excited when only one of the two driving components is present. As an additional illustration of a continuing resonance in the system, Fig. 3 shows a comparison between the velocity of the cnoidal wave in the lowest panel of Fig. 2 (blue line) and the phase velocity of the driving wave $V_1=\omega_{d1}/k_1$.

In order to avoid complex IST analysis of the autoresonant multiphase solutions (as in the example in the upper panel in Fig. 1), the next section describes our quasilinear Lagrangian approach to the problem, which is especially useful in studying the aforementioned autoresonant threshold phenomenon. We will see in the next Section that surprisingly, on the quasilinear level, the analysis reduces to evolution of much simpler, two fully decoupled one degrees of freedom dynamical problems in AA variables.

\section{Quasilinear problem and Whitham's averaging}
At this stage, we redefine $\eta =\varphi _{x}
$ via potential $\varphi $ satisfying 
\begin{equation}
\varphi _{xt}+6\varphi _{x}\varphi _{xx}+\varphi _{xxxx}=\epsilon _{1}\cos
\theta _{d1}+\epsilon _{2}\cos \theta _{d2},  \label{3}
\end{equation}%
We also notice that Eq (\ref{3}) can be viewed as Lagrange's
equation where the Lagrangian is
\begin{eqnarray}
L(\varphi ,\varphi _{x},\varphi _{t},\varphi _{xx}) &=& -\frac{1}{2}\varphi
_{x}\varphi _{t}-\varphi _{x}^{3}+\frac{1}{2}\varphi _{xx}^{2}
\nonumber\\&&- \varphi
(\epsilon _{1}\cos \theta _{d1}+\epsilon _{2}\cos \theta _{d2}).  \label{4}
\end{eqnarray}%
Prior to studying this system in its quasilinear limit, we write its linear solution
\begin{equation}
\varphi =a_{1}\cos \theta _{d1}+a_{2}\cos \theta _{d2},  \label{5a}
\end{equation}%
where%
\begin{equation}
a_{i}=\frac{\epsilon _{i}}{k_{i}(\omega _{di}-6\beta k_{i}+k_{i}^{3})},i=1,2.
\label{6}
\end{equation}%
Here, $\beta $ is the space-averaged $\left\langle \eta \right\rangle $.
Next, given the Lagrangian formulation above and the form of the linear
solution, we proceed to studying the general weakly nonlinear problem using
Whitham's averaged variational approach \cite{Whitham}. To this end, we write a
weakly nonlinear extension of Eq. \eqref{5a} 
\begin{eqnarray}
\varphi&=&\beta x-\int \gamma dt+a_{1}\cos \theta _{1}+a_{2}\cos \theta
_{2}+a_{11}\sin (2\theta _{1})+  \nonumber\\&&a_{22}\sin (2\theta _{2})+a_{12p}\sin (\theta
_{1}+\theta _{2})+\nonumber\\
&&a_{12m}\sin (\theta _{1}-\theta _{2}),  \label{7}
\end{eqnarray}%
where we view $a_{i}$ as small slowly varying first order amplitudes, while
the rest of the amplitudes are assumed to be of second order. The form of
the second order terms is chosen because of the second order nonlinearity. In addition, in Eq. \eqref{7}, we do not assume exact
phase-locking to the drives, but seek \textit{nearly} phase-locked state,
such that $\theta _{i}=k_{i}x-\int \omega _{i}(t)dt$, where $\omega
_{i}(t)\approx \omega _{di}(t)$ continuously. Consequently, in the driving
part of the Lagrangian, we replace
$\theta _{di}(t)=\theta _{i}(t)-\Phi _{i}(t),$  
where $\theta _{i}(t)$ are viewed as rapidly varying variables, while $\Phi
_{i}(t)$ are slow and bounded phase mismatches between the driven and
driving phases of the solution. Finally, the additional phase $\phi =\beta
x-\int \gamma dt$ in the ansatz (\ref{7}) is necessary, since $\varphi $ is
a potential and enters the Lagrangian through its space and time derivatives
only. Note that as before, $\beta =\left\langle \eta \right\rangle $.

At this stage, we proceed to Whitham's averaged variational approach \cite{Whitham} in our
problem, i.e., substitute anstz (\ref{7}) into Lagrangian (\ref{4}) and average it with respect to fast variables $\theta _{1}$ and $%
\theta _{2}$ between $0$ and $2\pi $. This yields the averaged Lagrangian%
\begin{equation}
\Lambda =\frac{1}{4}(\Lambda _{1}+\Lambda _{2}+\Lambda _{3}),  \label{10}
\end{equation}%
where%
\begin{eqnarray}
\Lambda _{1}&=&-4\beta \symbol{94}3-2\beta \gamma +k_{1}(\omega _{1}-6\beta
k_{1}+k_{1}^{3})a_{1}^{2}+\nonumber\\&&k_{2}(\omega _{2}-6\beta k_{2}+k_{2}^{3})a_{2}^{2},
\label{11}
\end{eqnarray}
    \begin{widetext}
\begin{eqnarray}
\Lambda _{2}
&=&6k_{1}^{3}a_{1}^{2}a_{11}+6k_{2}^{3}a_{2}^{2}a_{22}+6k_{1}k_{2}[a_{12m}(k_{2}-k_{1})+a_{12p}(k_{1}+k_{2})]a_{1}a_{2}+
4k_{1}(-6\beta k_{1}+4k_{1}^{3}+\omega
_{1})a_{11}^{2}+\nonumber \\
&&(k_{1}-k_{2})[(k_{1}-k_{2})^{3}+6\beta (k_{2}-k_{1})+\omega
_{1}-\omega _{2}]a_{12m}^{2}+4k_{2}(-6\beta k_{2}+4k_{2}^{3}+\omega
_{2})a_{22}^{2}+ (k_{1}+k_{2})[(k_{1}+k_{2})^{3}-\nonumber\\
&&6\beta (k_{1}+k_{2})+\omega
_{1}+\omega _{2}]a_{12p}^{2}  \label{11a}
\end{eqnarray}%
\end{widetext}
\begin{equation}
\Lambda _{3}=-2(\epsilon _{1}a_{1}\cos \Phi _{1}+\epsilon _{2}a_{2}\cos \Phi
_{2}).  \label{13}
\end{equation}%
Averaged Lagrangian $\Lambda $ depends on all $6$ slowly varying
amplitudes in ansatz (\ref{7}), on the auxiliary phase $\phi$ entering via
its space and time derivatives $\beta $ and $\gamma $, as well as on the two
phases $\theta _{1,2\text{ }}$of the solution via relation $\Phi _{i}=\theta
_{di}(t)-\theta _{i}(t)$ and their derivatives $k_{i}$ and $\omega _{i}$.
Since we have integrated over the fast phases, all the remaining objects in
the averaged Lagrangian are slow functions of time.
\section{Reduction to a two degrees of freedom dynamical problem}

This reduction is achieved by taking variations with respect to all
dependent variables. The first step is taking variation with respect to the
auxiliary phase $\phi$, yielding 
\begin{equation}
-\frac{\partial }{\partial t}\left( \frac{\partial \Lambda }{\partial \gamma 
}\right) =2\frac{\partial \beta }{\partial t}=0  \label{14}
\end{equation}%
and, thus, $\beta =\left\langle \eta \right\rangle $ remains constant given
by initial conditions. In all our simulations, we assume $\left\langle \eta \right\rangle=0$ initially and, therefore, in all developments bellow, use $\beta=0$. Next, we take variations with respect to first order
amplitudes $a_{1,2}$ to get
\begin{widetext}
\begin{eqnarray}
\frac{\partial\Lambda} {\partial a_{1} }&=&a_{1}k_{1}(k_{1}^{3}+\omega
_{1})+3[2a_{1}a_{11}k_{1}^{3}+a_{12p}a_{2}k_{1}k_{2}(k_{2}+k_{1})+a_{12m}a_{2}k_{1}k_{2}(k_{2}-k_{1})]-\epsilon _{1}\cos \Phi _{1}=0,
\label{15} \\
\frac{\partial \Lambda} {\partial a_{2}} &=&a_{2}k_{2}(k_{2}^{3}+\omega
_{2})+3[2a_{2}a_{22}k_{2}^{3}+a_{12p}a_{1}k_{1}k_{2}(k_{2}+k_{1})+a_{12m}a_{1}k_{1}k_{2}(k_{2}-k_{1})]-\epsilon _{2}\cos \Phi _{2}=0.
\label{16}
\end{eqnarray}
\end{widetext}
Similarly, variations with respect to second order amplitudes $%
a_{11},a_{22},a_{12p},$and $a_{12m}$ yield additional four algebraic
equations with solutions

\begin{equation}
a_{11} =-\frac{3a_{1}^{2}k_{1}^{2}}{4(4k_{1}^{3}+\omega _{1})}
\end{equation}
\begin{equation}
a_{22} =-\frac{3a_{2}^{2}k_{2}^{2}}{4(4k_{2}^{3}+\omega _{2})}
\label{18}
\end{equation}
\begin{equation}
a_{12p} =-\frac{3a_{1}a_{2}k_{1}k_{2}}{k_{1}^3+3k_{1}^2k_2+3k_{1}k_{2}^2+k_{2}^3+\omega _{1}+\omega _{2}}  \end{equation}
\begin{equation}
a_{12m} =\frac{3a_1a_2k_1k_2}{k_1^3-3k_1^2k_2+3k_1k_2^2-k_2^3+\omega_1-\omega_2} 
\end{equation}

We use the linear part of equations \eqref{15} and \eqref{16} to define two resonant frequencies.%
\begin{equation}
\omega _{0i}=-k_{i}^{3}.  \label{17a}
\end{equation}%
In the following discussions, being interested in the passage through these
linear resonances, we write $\omega _{di}=\omega _{0i}+\alpha _{i}t$ and,
assuming proximity to the resonances near $t=0$, substitute $\omega _{0i}$
instead of $\omega _{i}$ in the second order amplitudes.
This yields%
\begin{eqnarray}
a_{11} &=&-\frac{a_{1}^{2}}{4k_{1}},  \nonumber \\
a_{22} &=&-\frac{a_{2}^{2}}{4k_{2}},  \label{19} \\
a_{12p} &=&-\frac{a_{1}a_{2}}{k_{1}+k_{2}},  \nonumber \\
a_{12m} &=&-\frac{a_{1}a_{2}}{k_{1}-k_{2}}.  \nonumber
\end{eqnarray}%
Next, we return to Eqs. \eqref{15} and \eqref{16}, which after using Eqs. \eqref{17a} and \eqref{19} become

\begin{equation}
\omega _{i}-\omega _{0i}-\frac{3}{2}k_{i}a_{i}^{2}-\frac{\epsilon _{i}}{%
a_{i}k_{i}}\cos \Phi _{i}=0,  \label{21}
\end{equation}%
By using the definitions, we can write
\begin{equation}
\omega _{i}-\omega _{0i}=(\omega _{i}-\omega _{di})+(\omega _{di}-\omega
_{0i})=-\frac{d\Phi _{i}}{dt}+\alpha _{i}t.  \label{22}
\end{equation}%
Then Eqs. (\ref{21}) yield equations for $\Phi _{i}$:%
\begin{equation}
\frac{d\Phi _{i}}{dt}=\alpha _{i}t-\frac{3}{2}k_{i}a_{i}^{2}-\frac{\epsilon
_{i}}{a_{i}k_{i}}\cos \Phi _{i}.  \label{24}
\end{equation}%
Finally, we write Lagrange's equations for the angle variables $\theta _{i}$ 
\begin{equation}
\frac{\partial }{\partial x}\left( \frac{\partial \Lambda }{\partial k_{i}}%
\right) -\frac{\partial }{\partial t}\left( \frac{\partial \Lambda }{%
\partial \omega _{i}}\right) =\frac{\partial \Lambda }{\partial \Phi _{i}}
\label{25}
\end{equation}%
yielding lowest significant order differential equations for the
amplitudes%
\begin{equation}
\frac{da_{i}}{dt}=-\frac{\epsilon_{i}}{k_{i}}\sin \Phi _{i}.  \label{27}
\end{equation}%
Equations (\ref{24}) and (\ref{27}) comprise a complete set of dynamical
equations for the two pairs of variables $(\theta _{1},a_{1})$ and $(\theta
_{2},a_{2})$ in our problem. Remarkably, these two degrees of freedom are
fully separated in the weakly nonlinear approximation, which to our
knowledge is characteristic to the KdV system only. For example, long
wavelength ion acoustic wave in plasmas are frequently modeled by the KdV
equation. However, as shown by Munirov et al. recently \cite{Munirov}, using the
full set of plasma fluid equations that describe ion acoustic waves, the two
degrees of freedom for two-phase waves in this case are not separable,
meaning the impossibility of reduction to the KdV model. The separation of
variables simplifies the problem of the thresholds for capture in
resonance. Indeed, for each of the two degrees of freedom one uses the simple
single degree of freedom theory in the problem \cite{Wiki} to get two independent thresholds on
the driving amplitudes, i.e., 
\begin{equation}
\epsilon _{th}^{i}=0.41(2k_{i}/3)^{1/2}\alpha _{i}^{3/4}.
\label{28}
\end{equation}%
This separability was tested by finding the threshold numerically in \cite{Lazar2} and now we have an analytic justification. 
\begin{figure}[tbh]
\includegraphics[scale=1.2]{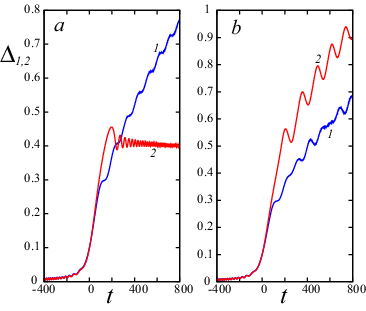}
\caption{The widths of the IST spectrum gaps versus time. Left panel:  $\epsilon_1=1.5\epsilon^1_{th}$, $\epsilon_2=0.99\epsilon^2_{th}$,
Right panel: $\epsilon_1=1.5\epsilon^1_{th}$, $\epsilon_2=1.01\epsilon^2_{th}$. In both panels, $\alpha_{1,2}=0.001$.}
\label{FIG4}
\end{figure}

\section{IST confirmaion}
The main conclusion of our theory above is that the two degrees of freedom associated with the two phases of the wave are fully separated in the weakly nonlinear
limit. In this section, we illustrate numerically this unique property of the KdV system and the associated autoresonant control at finite amplitudes via the IST approach of Ref.  \cite{Lazar2}. Consider again the KdV equation driven by two chirped frequency drives with $k_{1,2}=1,2$ and $\alpha_{1,2}=0.0001$. We set the driving amplitude of the first driving component $\epsilon_1=1.5\epsilon_{th}^1$ and vary the second driving amplitude to find the associated threshold. Thus, we discuss the threshold phenomenon for the second mode in the background of the first autoresonant mode. The IST theory of Ref. \cite{Lazar2} is based on analyses of the IST spectrum of the driven KdV equation. The two-phase driving opens two gaps in this spectrum, indicating the formation of two phases in the solution. The time evolution of the widths of these gaps ($\Delta_{1,2}$) is shown in Fig. 4. The left panel of the figure shows the case where the first degree of freedom is phase locked to its drive, while the second degree of freedom dephases at later times. This is the case below the threshold as $\epsilon_2=0.99\epsilon_{th}^2$. In contrast, the right panel shows the case $\epsilon_2=1.01\epsilon_{th}^2$, and one can see the continued growth of the width of the gap, indicating the autoresonant two-phase evolution. Similar simulations allow high-accuracy estimates of the threshold. Figure 5 shows (solid red circles) the threshold values $\epsilon_{th}^2$ found numerically using this procedure as a function of $\alpha_2$ in the log-log plot. The
dashed line in the figure shows the theoretical threshold, Eq. \eqref{28}. One can see a good agreement between numerical and theoretical results up to $\alpha_2\approx0.1$. These calculations also demonstrate the separation of the two degrees of freedom in the KdV equation even for rather large amplitudes of drivings. 
\begin{figure}[tbh]
\includegraphics{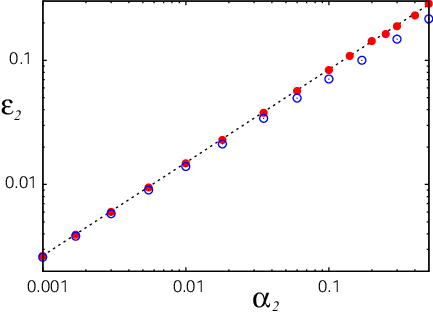}
\caption{The threshold for autoresonant excitation of mode $k=2$. Dashed line: theory, \eqref{28}, full red circles: two-phase wave, open circles: four-phase wave.}
\label{Fig. 5}
\end{figure}

\begin{figure}[tbh]
\includegraphics[scale=1.2]{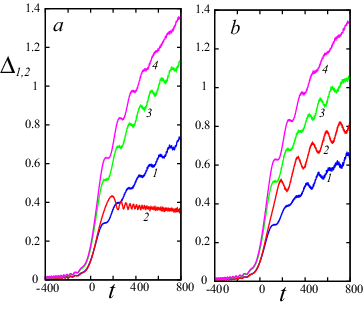}
\caption{The evolution of the IST spectrum gaps of a four-phase wave with driving parameters $\epsilon_1=1.5\epsilon^1_{th}$,
$\epsilon_3=1.4\epsilon^3_{th}$,
$\epsilon_4=1.3\epsilon^4_{th}$
$\alpha_4=\alpha_3=\alpha_2=\alpha_1=0.001$.
The left panel: $\epsilon_2=0.97\epsilon^2_{th}$. The right panel: $\epsilon_2=1.01\epsilon^2_{th}$}
\label{Fig. 6}
\end{figure}

\begin{figure}
   \includegraphics[scale=0.61, left]{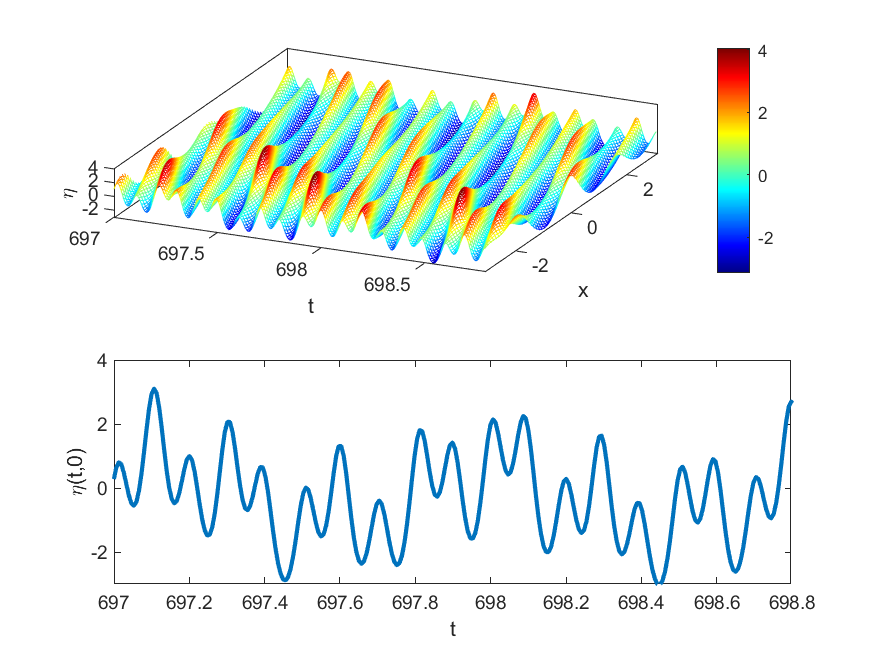}
    \caption{The autoresonant 4-phase wave. Upper panel: the evolution in space-time, Lower panel: the time dependence of the wave at x=0.}
    \label{Fig. 7}
    \end{figure}

Finally, we test the threshold for the same $k=2$ mode when in addition to $k=1$ mode, two new phase-locked modes are added with $k=3$ and $k=4$. Figure 6 with the additional two modes is similar to Fig. 4 and allows one to find the autoresonant threshold on the second driving amplitude by a similar procedure. In this calculation, we
set $\alpha_1=\alpha_2=\alpha_3=\alpha_4$ and show by blue circles in Fig. 5
the numerical estimates for the threshold of mode
k = 2 for the four-mode driving showing that Eq. \eqref{28} remains valid at sufficiently small amplitudes
of the driving. Thus, we have an indication that the separation of
the degrees of freedom takes place for multi-mode drivings as well. This separation of the degrees of freedom means that all modes at the weakly nonlinear stage are excited independently, and therefore, we can easily form very complex multiphase patterns which remain in autoresonance under full control via the driving. An example of autoresonant excitation of a complex pattern by four-mode driving is illustrated
in the upper panel of Fig. 7, while the lower panel shows the evolution of $\eta(t,0)$ at the center $x = 0$ of the computation interval. Despite the complexity of this solution, the evolution of its IST spectrum shown in Fig. 8 is rather simple. \begin{figure}[tbh]
\includegraphics[scale=0.85,left]{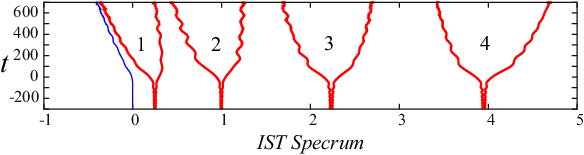}
\caption{The opening of 4-phase IST spectrum gaps in autoresonant evolution}
\label{Fig. 8}
\end{figure}
After
crossing the resonance at $t = 0$ four gaps are opened that correspond to four different wave numbers $k_i$ of the driving components. The time dependence of the widths of the gaps is controlled by the corresponding chirp rates $\alpha_i$ of the drivings, if the threshold conditions for all modes are satisfied.

\section{conclusions}
We have studied autoresonant (continuously phase-locked) excitation and control of two-phase KdV waves. The process involved driving the system by two independent, small amplitude, chirped frequency driving waves. We illustrated these excitations in computer simulations (see Figs. 1-3). 

One of the important aspects of autoresonant excitations is the existence of a threshold (see Eq. \eqref{28}) on the driving amplitudes for entering the phase-locked regime. We have developed the theory of these thresholds using a weakly nonlinear theory. This approach significantly simplifies the analysis for multiphase waves, which are usually described by more complex inverse scattering theory (IST) \cite{Novikov}. The weakly nonlinear theory was developed in Sec. III through Whitham's averaged Lagrangian approach. It allowed a reduction of the problem to a dynamical two degrees of freedom system in action-angle variables. Furthermore, we have found that the two degrees of freedom are fully separated (see Eqs. \eqref{24} and \eqref{27}), significantly simplifying the derivation of the autoresonant thresholds in the problem. We have also confirmed this separation of the two degrees of freedom in Sec. V via the IST theory. 

In addition, we have excited much more complex autoresonant 4-phase waves in computer simulations (see Fig. 7). Despite the complexity of these solutions in space-time, their IST spectrum is very simple (see Fig. 8). Studying the threshold for autoresonant excitations in these simulations suggested that again all degrees of freedom are separated on the weakly nonlinear level. 

Finally, the addition of dissipation in the multi-phase KdV problem and the application of similar ideas to other integrable nonlinear wave equations seem to comprise interesting directions for future research. 
\section*{Acknowledgement}
A.G.S. was supported by the Ministry of Science and Higher Education of the Russian Federation (theme "Quantum", Grant No. 122021000038-7).

\end{document}